\documentclass[aps,prb,twocolumn,a4paper,superscriptaddress]{revtex4-1}
\usepackage{graphicx}
\usepackage{color}

\newcommand{\etal}{\textit{et al.\/}}

\begin{document}

\title{Electronic structure and magnetic properties of Mn, Co, and Ni-substitution of Fe in Fe$_4$N}
\author{Patrizia Monachesi}
\affiliation{Dipartimento di Scienze Fisiche e Chimiche,Universit\'a dell'Aquila, Via Vetoio, I-67010 L'Aquila, Italy}
\author{Torbj\"orn Bj\"orkman}
\affiliation{COMP/Department of Applied Physics, Aalto University, FI-00076 Aalto, Finland}
\author{Thomas Gasche}
\affiliation{CFMC-UL, Dep. Fisica, Faculdade de Ci\^encias, Universidade de Lisboa and CINAMIL, Laboratorio de Fisica, Military Academy, Lisbon, Portugal}
\author{Olle Eriksson}
\affiliation{Department of Physics and Astronomy, Uppsala University, Box 516,
  751\,20 Uppsala, Sweden}

\date{\today}

\begin{abstract}
The magnetic properties of Mn, Co and Ni substituted Fe$_4$N are calculated from first principles theory. It is found that the generalized gradient approximation reproduces with good accuracy the magnetic moment and equilibrium volume for the parent Fe$_4$N structure, with the atomic moment largest for the Fe atom furthest away from the N atom (Fe I site), approaching a value of 3 $\mu_B$/atom, whereas the Fe atom closer to the N atom (Fe II site) has a moment closer to that of bcc Fe.
Substitution of Fe for Mn, Co or Ni, shows an intricate behavior in which the Mn substitution clearly favors the Fe II site,  Ni favors substitution on the Fe I site and Co shows no strong preference for either lattice site. Ni and Co substitution results in a ferromagnetic coupling to the Fe atoms, whereas Mn couples antiferromagnetically on the Fe II site and ferromagnetically on the Fe I site. For all types of doping, the total magnetic moment is enhanced compared with Fe$_4$N only in the energetically very unfavorable case of Mn doping at the Fe I site.
\end{abstract}

\pacs{75.50.-y,75.30.Cr,71.20.Eh,75.50.Bb}


\maketitle

\section{Introduction}
Magnetoresistive devices, electromagnetic motors, transformers and generators have their performance dictated by the size of the magnetic moment and the strength of the coercive field.\cite{ekkes,coey,mohn:magnetism,kuebler:itinerantmagnetism,thompson00ibmjresdev44:311} This holds for bulk magnetic materials as well as for nano-sized objects. Applications in high density recording media and sensors, e.g., rely on nanostructures derived from parent bulk compounds and alloys, which should have a high value of the unit cell magnetization. Technology in different areas, calls for magnetic materials with improved properties. As concerns the saturation moment, the room temperature value 2.45 $\mu_B$/atom for a bcc Fe-Co alloy, as described by the 
maximum of the Slater-Pauling curve,\cite{pauling,weiss29adp12:279,stoner:magnetism} has been difficult to exceed, although recent efforts have suggested promising nano-laminates.\cite{sanyal,caroline}  

Other efforts in finding materials with enhanced saturation moments were made by Bergman \etal \cite{bergman04prb70:174446}, who found from first principles theory, that a high saturation magnetization should be possible by a close packing of small Fe clusters in a Co matrix. Experimental works\cite{baker02jmmm247:19,binns01sscrep44:1} exploring this idea, were unfortunately not conclusive, whether or not a saturation moment larger than that of the Slater-Pauling maximum was reached.

Another group of promising  elements  have been considered:  the 3d transition metal nitrides.
Fe, Ni and Co nitrides are conducting ferromagnets with high mechanical resistance, low coercivity and potentially large saturation moments.     
Some experimental studies have indeed proposed Fe$_4$N,  crystallizing in the perovskite structure, as a candidate material \cite{kim72apl20:492}
for which the saturation moment per atom is larger than the Slater-Pauling 
maximum, but other reports contradict this finding.\cite{sakuma96jap79:5570} 
Moreover, 
Patwari and Victora \cite{patwari01prb64:214417} , based on calculations relying on density functional  
theory, showed that the addition of Mn to Fe$_4$N increases the unit cell volume 
and the magnetization. However, the magnetic moment per atom 
was found not to exceed that of the Fe-Co alloys.

\begin{figure}
\includegraphics*[width=9cm]{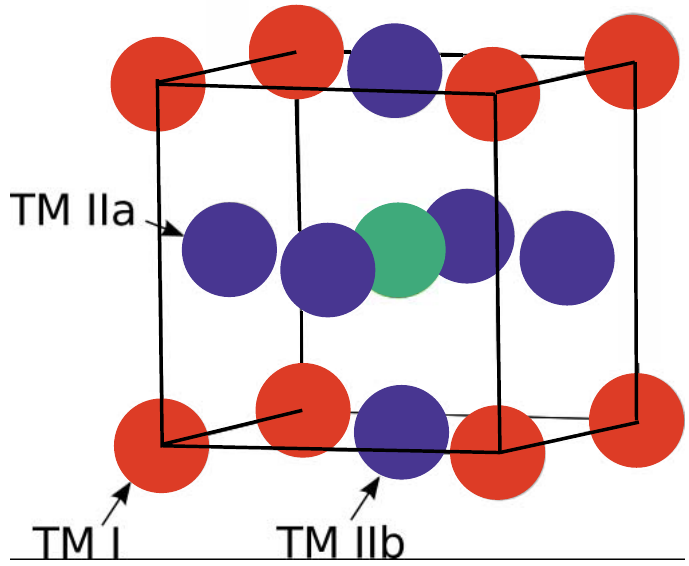}
\caption{(Color online) Perovskite structure of Fe$_4$N with the indication of the twofold positions of Fe II ions and the position of the N atom, in the middle of the unit cell.}
\label{fig1}
\end{figure}

A recent review article\cite{eitel} comparing the outcomes of  several  ab-initio calculations of Fe$_4$N among them and  also with the few available experimental data suggest, however, that a saturation moment larger than the Slater-Pauling maximum is indeed possible. The different conclusions concerning the possible exploitation  of the magnetic properties of Fe$_4$N highlight the complexity of the electronic structure  of this magnetic material to which the present investigation addresses. So far one can conclude that in  Fe$_4$N  there are saturated (high spin)  and unsaturated (low spin) moments of the Fe atoms at the corners (Fe I) and face centers (Fe II) of the perovskite structure, respectively. This is found in all calculations.\cite{eitel}
Fig. \ref{fig1} shows  the perovskite structure of  Fe$_4$N  with the  five  atomic positions  in the conventional unit cell,  labelled  as  in Eitel et al \cite{eitel}. Atoms at positions I  are a factor  $\sqrt{3}$  further away from N than those at positions II: a sound reason for the  atomic moment to reach saturation at this position. On the other hand, atoms at  positions II, further distinguishable in the presence of spin-orbit interaction according to the direction of their 4-fold rotation axis relative to the moments, are easily hybridized with neighboring nitrogen to an extent that is strongly dependent on volume and chemical coordination.  

The calculated values of the moment of the  Fe I and Fe II site may differ by up to 20\%  depending on the calculation method and on the choice of exchange-correlation potential in the Hamiltonian\cite{eitel}. The complex magnetic behavior  of  the Fe atoms in Fe$_4$N is  not  restricted to   this compound. It is in fact observed  in Co$_4$N\cite{matar1} and  in  the Mn$_4$X series (with X=N,C,B, or Be), where  the magnetic cell  is found to be ferrimagnetic at an increased volume, with three different values of the Mn moment at position I, IIa and IIb\cite{patwari01prb64:214417}.

The possibility to enhance the magnetic performances of Fe$_4$N by alloying has not been fully explored and is the aim of this study. Scattered investigations along this line of thoughts are available for ordered compounds TMFe$_3$N (TM = Ti, Cr, Co, Ni, Pd), but are all limited to calculations for the most symmetrical situation, with TM atom substituting the Fe I at the corner position\cite{mohn-schwarz,santos}. In  the first principles investigation presented here, we consider the substitution of TM (Ni, Co, Mn)  atoms on both Fe sites, and compare the phase stability and magnetic properties of the different compounds systematically  within the same computational scheme.   

The paper is organized as follows. The next section (Section II) describes the calculations methods, giving the essential input parameters and approximations adopted to achieve the desired numerical convergence.  In section III we give the results for the reference compound Fe$_4$N whose magnetic properties are obtained by two different methods each  with  either Local Spin Density (LDA) and Generalized Gradient (GGA) approximations, with or without inclusion of orbital polarization.   
We then describe the magnetic properties and structural stability of the substituted compounds. In  the concluding section we try to draw a picture of the magnetic properties of TM nitrides and compare with previous results. 

\section{Method and calculations}

To compare the electronic structure and magnetic moments of several ordered  TMFe$_3$N systems  relative to Fe$_4$N, we first  fix a quantitative, reference picture of   Fe$_4$N  based on our simulation scheme, since, as discussed above,  total  energy and magnetic moments depend on the exchange-correlation (XC) scheme and on the unit cell volume.

All DFT calculations, self-consistent and {\it ab initio}, were performed  in LDA and GGA of the exchange-correlation potential.  The calculations were performed  by a fully-relativistic implementation of the full-potential linear muffin tin orbitals (FP-LMTO) method. \cite{willscooper,wills:fp-lmto,wills:fp-lmto2}. A so-called triple basis was used to ensure a good convergence of the wave functions allowing us to compare total energies with an accuracy of mRy/f.u. and magnetic moments within one hundredth of a Bohr magneton.  We have also performed calculations by the FPLO method \cite{FPLO,FPLO2}, scalar-relativistic and with the same XC approximation,  to estimate  the possible quantitative spread in the results. 
We used a 16x16x16 Monkhorst-Pack mesh in FP-LMTO calculations and a cubic 10x10x10 mesh in FPLO calculations.
\section{Results }
\subsection{Fe$_4$N }

\begin{figure}
\includegraphics[width=9cm]{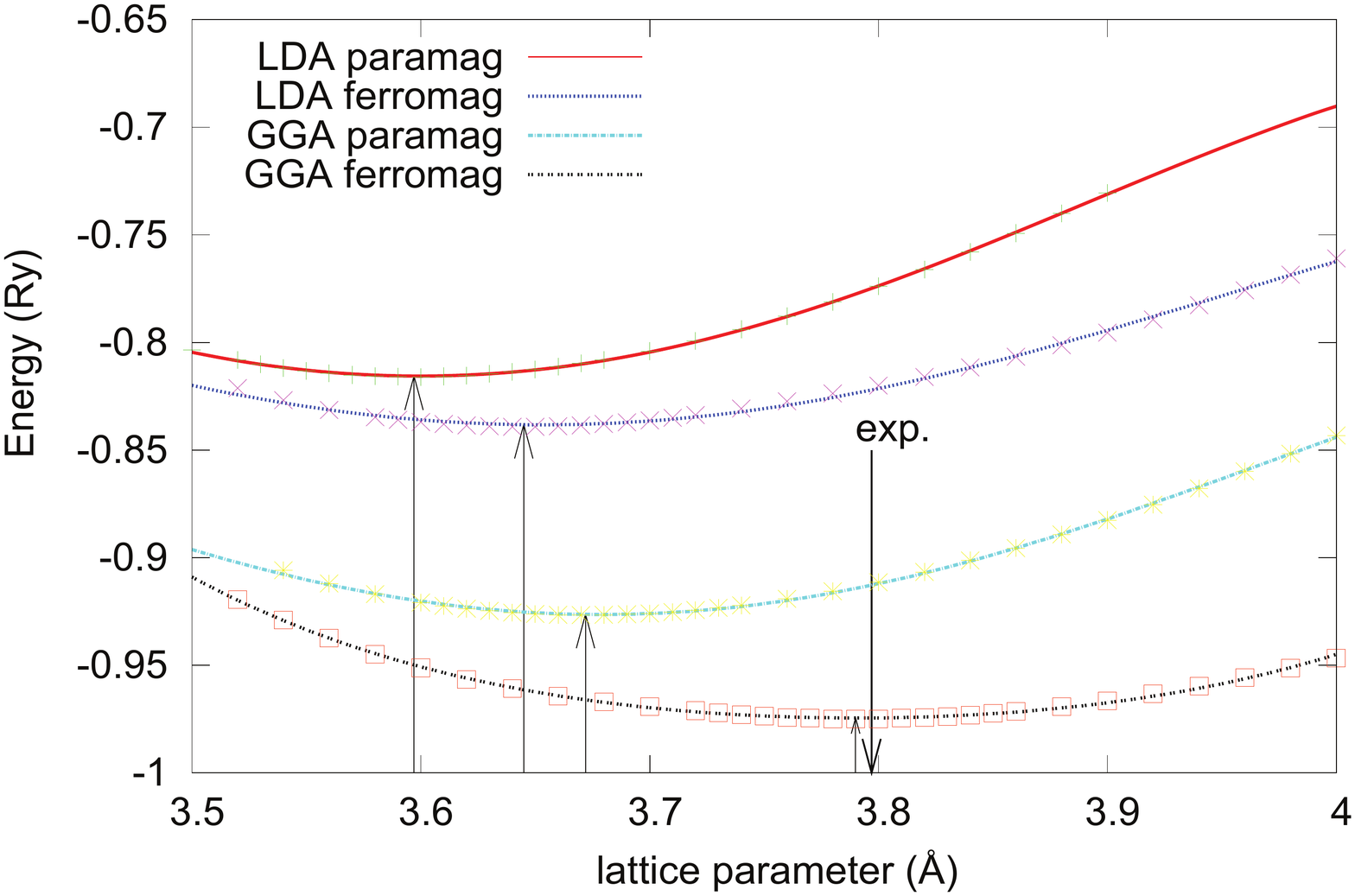}
\caption{ (Color online) Calculations for equilibrium lattice parameter  of Fe$_4$N using LDA and GGA and the FPLO method for paramagnetic and ferromagnetic  configurations. The energies of spin-polarized GGA results are shown by open squares.  Minima in each curve are  indicated by vertical arrows. The experimental equilibrium lattice parameter a = 3.796 \AA{} is  indicated by a downward arrow. }
\label{fig2}
\end{figure}

A first set of reference calculations were done for Fe$_4$N, by calculating the total energy curve and magnetic moment as a function of the lattice constant since  volume plays a critical role in the magnetism of this compound. In Fig.\ref{fig2} we show results for both spin-polarized and paramagnetic, i.e. spin degenerate, calculations, using LDA and GGA functionals for comparison. The curves in Fig.\ref{fig2}  show that spin-polarization occurs spontaneously in the calculations, since the corresponding total energy is lower than for the spin-degenerate state, both for LDA and GGA functionals. Moreover, as also pointed out in previous papers\cite{matar1}, GGA  calculations  seem more appropriate for these compounds since the total energy of the  ferromagnetic phase 
has its   minimum at a$_{GGA}$ = 3.789 {\AA}, very  close to the experimental lattice parameter  $a$ = 3.797 {\AA} found by Frazer\cite{frazer}  and  $a$ =  3.790 {\AA} by Jacobs {\it  et al}\cite{exp0}. Also worth noticing is  the large difference 
between the GGA and LDA equilibrium lattice parameters. 

\begin{figure}
\includegraphics[width=9cm]{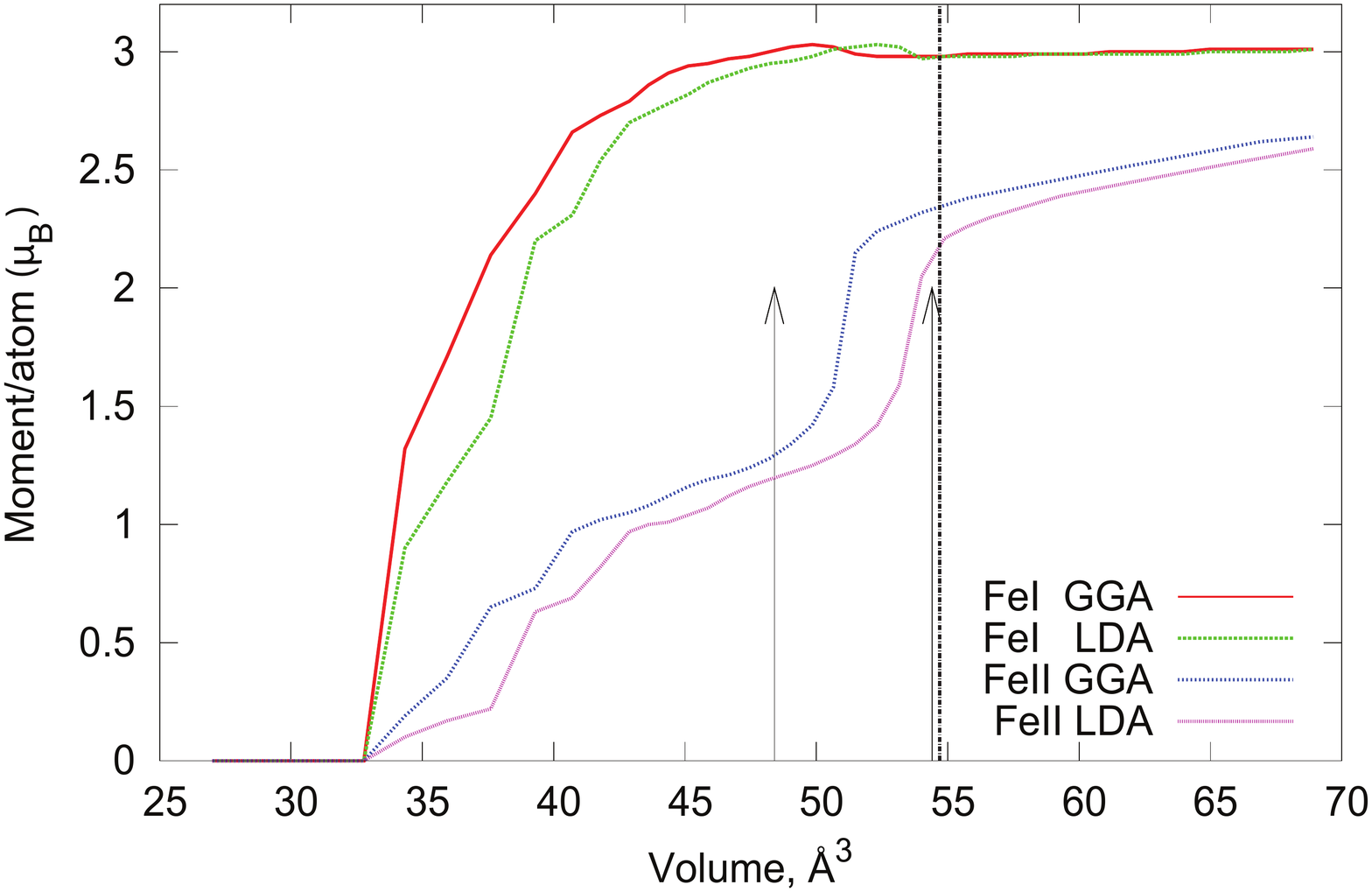}
\caption{(Color online)  Magnetic moments of Fe at different sites as function of volume in Fe$_4$N. The experimental volume is marked by a vertical line whereas the calculated equilibrium volumes  are indicated by two vertical arrows corresponding to GGA and LDA calculations, where the GGA arrow is closest to the experimental volume. }
\label{fig3}
\end{figure} 

Fig.\ref{fig3} shows the non-trivial behavior of the magnetic  moment in ferromagnetic Fe$_4$N as a function of volume.  
Both LDA and GGA  curves confirm that  the Fe I moment already reaches saturation at a volume smaller that the experimental equilibrium volume.  In contrast, the Fe II moment undergoes a step-wise increase of roughly 1 $\mu_B$/atom at a volume between the LDA and GGA  equilibrium volumes. Moreover,  at  the LDA equilibrium volume (left side arrow in Fig. 3) the  Fe II moment state has not reached the steady low spin behavior at variance with the moment obtained at and beyond the GGA equilibrium volume (right side  arrow in Fig. 3). Hence we conclude that GGA describes better the magnetic and cohesive properties of these compounds, as also pointed out in references 18-20.
From the GGA calculation, the total magnetic moment at the experimental lattice constant is 9.84$\mu_B$/f.u. to which contribute the magnetic moments of Fe I and Fe II  (see Fig.\ref{fig1})  by   2.91 and 2.31 $\mu_B$/atom, respectively. 
Our FPLO GGA calculations result in a total magnetic moment at the equilibrium  lattice constant of 9.93$\mu_B$/f.u., i.e. a value close to the FP-LMTO result.
Our calculated values fit quite well in the range of  results obtained by many different simulations, as well as with the few experimental results available (see Table 1 of  Eitel{\it et al}\cite{eitel}).

\begin{figure}
\includegraphics[width=9cm]{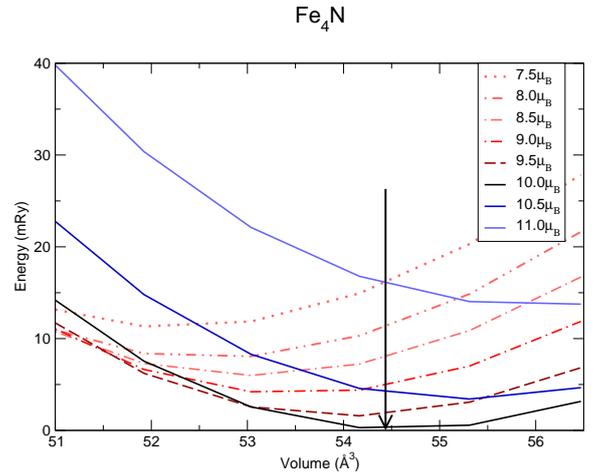}
\caption{(Color online) Total  energy vs unit cell volume for different  magnetic moments, calculated using the fixed spin-moment technique.}
\label{fig4}
\end{figure}

We  test further the relationship between volume and magnetism  of Fe$_4$N with volumes given in the range between between the  LDA and GGA results (see Fig.\ref{fig3}),   by  plotting in Fig.\ref{fig4}      
the total energy vs unit cell volume for different fixed values of the unit cell moment, in the range  7.5 - 11.0 $\mu_B$/f.u. in steps of 0.5 $\mu_B$/f.u. These results are obtained  with the fixed spin technique in the FP-LMTO calculation in GGA. Note that the calculations with larger spin moment have larger equilibrium volume, due to the magneto-volume effect.  It is also seen that the curve with a fixed moment of 10.0  $\mu_B$/f.u.has the lowest energy of all curves shown in Fig.\ref{fig4}, since this is the magnetic moment closest to the moment of the ground state configuration (9.84 $\mu_B$/f.u.), as discussed above.  Fig.\ref{fig4} also shows that  increasing the saturation moment from the ground state value, to e.g. 10.5 $\mu_B$/f.u. brings about  an increase of $\sim$ 5 mRy/f.u. which in this  context  is a rather large value. Furthermore, considering that the nitrogen spin contribution is negligible, this plot is in complete agreement with the findings in Fig.\ref{fig3} confirming the critical  dependence of the moments upon  the unit cell volume, on one side, and pointing out  the dramatically  different results  given  by GGA vs. LDA for  the equilibrium volume and magnetic moment.      

\subsection{Ni, Co, Mn substitution}

As pointed out in the introduction,  substitution of one Fe atom by other $3d$ transition metals  has been  investigated only to a limited amount  experimentally. Moreover, the theoretical calculations  always address substitution of  the Fe I atom, except for a work by Patwari and Victora\cite{patwari01prb64:214417} who explored the magnetic properties of  Mn-Fe nitrides with different compositions.  

Here we systematically substitute Ni, Co and Mn at  each of the two Fe positions shown in Fig.\ref{fig1} forming compounds TMFe$_3$N (where TM is Mn, Co or Ni), and we calculate the magnetic properties using the GGA scheme, which was found to be more appropriate, as discussed above. Calculations are done  at the TMFe$_3$N's   equilibrium volumes reported in Table I. We find that substituting at the Fe-I or Fe II sites (a- or b-position in Fig.1) by Ni and Co always gives rise to a ferromagnetic alignment, in contrast to  Mn substitution which aligns ferromagnetically only for substitution on the Fe I site and induces ferrimagnetic order for substitution at Fe II (discussed below).

\begin{table}[ht]
\caption{Equilibrium volumes ($\AA^3$) of  TMFe$_3$N  compounds calculated by FP-LMTO. For Mn both ferromagnetic (FM) and antiferromagnetic (AFM)  states are reported. Volumes obtained by   FPLO calculations agree within 0.4\%.} 
\centering 
\begin{tabular}{c | c | c | c} 
\hline\hline 
Substituted Fe site &	Mn (FM)/Mn (AFM) &	Co (FM)	&  Ni (FM) \\
\hline
  TM@FeI & 55.79 /  53.59  &          53.66  &  53.80  \\ 
  TM@Fe II & 54.73 / 54.08  &           53.19  &          53.80  \\ [1ex] 
\hline 
\end{tabular}
\label{table:nonlin} 
\end{table}

\begin{figure}
\includegraphics[width=9cm]{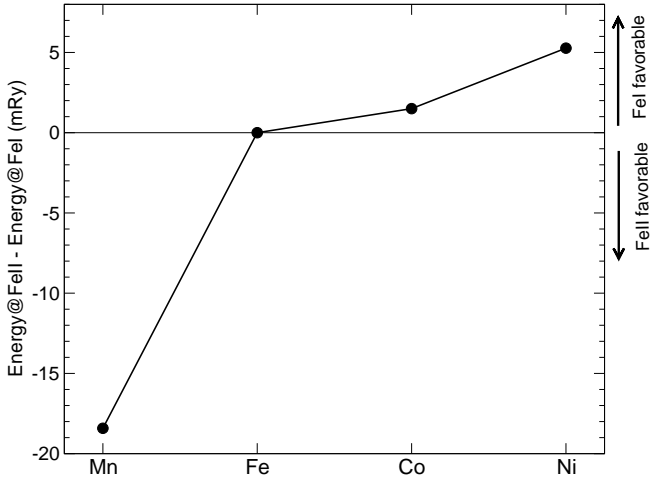}
\caption{Total energy difference E$_{II}$-E$_{I}$  when TM atoms (Mn,  Co, or Ni) substitute at Fe I position and Fe II position in TMFe$_3$N. The stability of TM atoms occupying the Fe I or Fe II site is indicated on the right side of the plot.}
\label{fig5}
\end{figure}

The stability of the TM-substituted structure for different TM elements, either in Fe I or Fe II positions, is illustrated in Fig.\ref{fig5}. Here we plot the difference in total energy of the TMFe$_3$N compounds with TM atoms substituting at the Fe I or Fe II site. From this figure it is seen that Mn has a strong preference to occupy the Fe II site, whereas Co only has a weak preference to occupy the Fe I site. This points to the possibility to stabilize a random alloy of Co atoms on the Fe I and Fe II site, if configurational entropy is considered. Finally, Ni substitution is found to clearly favor the Fe I site. 

\begin{figure}
\includegraphics[width=9cm]{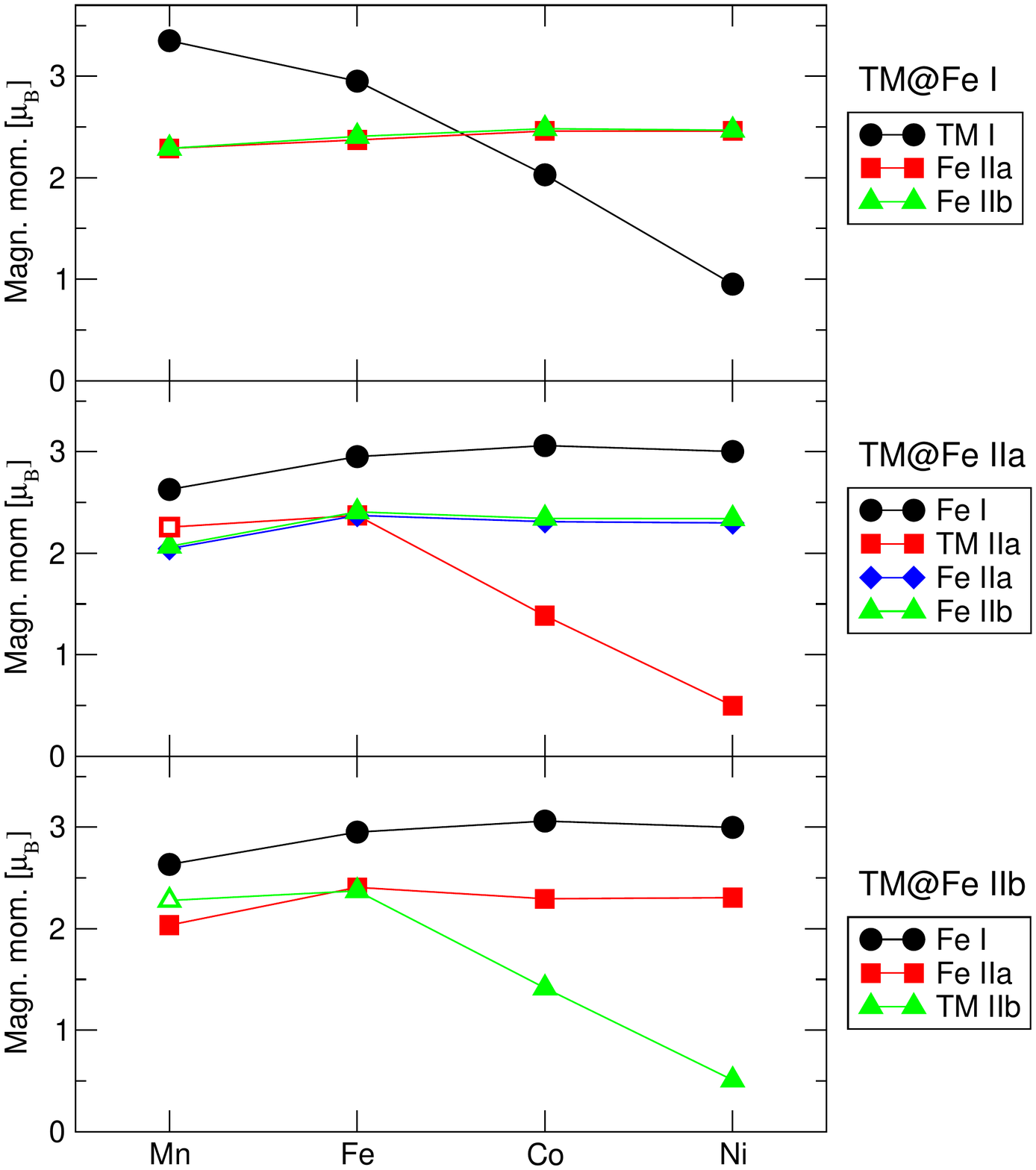}
\caption{(Color online) Magnetic moments of transition metal atoms in the different lattice sites of the  Fe$_4$N unit cell. Open symbols for Mn indicate an antiferromagnetic coupling between this atomic moment and the Fe moments.}
\label{fig6}
\end{figure}

The results presented in Fig.\ref{fig5} are further analyzed by calculating the magnetic moments of all atoms at each position in the different compounds, as shown in Fig.\ref{fig6}.  Here it can be seen that the moments of  any atom  sitting at Fe I site  exceed, by at least 0.5  $\mu_B$/atom , those  at  Fe II site.

The substitution of Mn on the Fe I site gives a clear enhancement of the total moment over Fe$_4$N, due to the even higher moment of  Mn than Fe and the ferromagnetic coupling. Unfortunately, this configuration is energetically very unfavorable, so it appears most unlikely that a compound with the composition MnFe$_3$N could be synthesized in this form. When doping on the Fe I site, the equilibrium volume is markedly larger than when doping on the Fe II site, which also accounts for a small enhancement of the overall moment in that case, as found in calculations not shown here. Mn doping at site II, occurs for an antiferromagnetic coupling between Mn and Fe atoms at site I. Both the Fe and Mn moments are rather large (see Fig.\ref{fig6}), but the antiferromagnetic coupling of them forces the net moment of the unit cell to unfortunately not be  enhanced compared to Fe$_4$N. Substitution of Co or  Ni, on either of the Fe I or Fe II sites, results in a ferromagnetic coupling, albeit with a  magnetic moment per f.u. which is lower than that of  Fe$_4$N, since the atomic Co and Ni moment is smaller than that of Fe (Fig.\ref{fig6}). The orbital moments  were found  to be small, the largest contribution  amounting to  0.07 $\mu_B$ (always pointing in the same direction as the spin moments as expected for  more than  half-filled shells). Their effect reflects  the almost imperceivable difference between the green and blue curves in the middle panel of Fig.\ref{fig6}.

\subsection{Influence of N vacancies}
We also considered effects due to vacancies in our calculations. The vacancy calculation was done by removing one N atom in a 3x3x3 supercell and allowing the atoms closest to the vacancy to relax. The results were similar when carried out in a 2x2x2 supercell, showing that supercell convergence was achieved with respect to the local magnetic moments. The relaxations of atoms beyond the 6 nearest Fe atoms were very small and appear to have no discernible effect on the magnetism, so relaxing only the four nearest neighbor shells seems well justified. The effect of the defect was to increase the on-site moment of the Fe atoms closest to the vacancy by about 0.2 $\mu_B$ per atom and the changes on all other atoms, compared to the pristine system, were smaller than 0.01 $\mu_B$. However, despite increase of the on-site moments close to the vacancy, the total moment per Fe atom remains at 2.46 $\mu_B$/Fe, the same as for the pristine system. The decrease of the magnetic moment in other parts of the system is spread over a large portion of the unit cell without any clearly identifiable primary source. 

\subsection{Calculations at the substitutional site in the Virtual Crystal Approximation}
VCA calculations were then carried out for the substitutional atom at both site I and site II, thus allowing an artificial change of the atomic number (and number of electrons) to non-integer values from Z=25 to Z=28 (i.e. from Mn to Ni). For each VCA value the lattice parameter was found by energy minimization for
both ferromagnetic and antiferromagnetic coupling of the substitutional atom.

Substituting one VCA atom for an Fe atom at site I results in a ferromagnetic solution for the whole range of atomic numbers, from Z=25 to Z=28. The magnetic moment on the VCA atom has a highest value of almost 3.7 $\mu_B$ for Z=25.0 and declines almost linearly to 0.7 $\mu_B$ for  Z=28.0. Simultaneously, the  moments  of  Fe II atom rise from 2.2 $\mu_B$ ( at Z=25.0)  to 2.5 $\mu_B$ (at  Z=28.0), a trend already found  in Fig.\ref{fig6}. The combination of the slow increase in the site II moments and the rapid decrease in the site I moment results in a total moment that starts at 10.2 (Z=25.0) , rises slightly to 10.3 $\mu_B$/f.u. at Z=25.1 / Z= 25.2 before declining steadily to 8 $\mu_B$ at Z=28.0. Forcing an antiferromagnetic coupling on the VCA atom results in an energy that is between 50-60 mRy higher than for the ferromagnetic coupling.

Substituting one VCA atom for an Fe atom at site II results in a ferromagnetic coupling down to Z=25.2 (almost Mn). Antiferromagnetic coupling has a lower energy only for the cases of  VCA atoms with Z=25.0 and 25.1. The highest magnetic moment is found for Z=25.5,  10.1 $\mu_B$/f.u.. At Z=26 the antiferromagnetic solution is found to be only 3 mRy higher in energy than the ferromagnetic one and, as noted above, crosses over the ferromagnetic energy
curve below Z=25.2.  

A comparison of the two groups of calculations shows that for Z $<$ 26 the preferred site is II and for Z $>$ 26.0 the preferred site passes over to site I, as already noted in the previous section on substitutions with Mn, Co and Ni. 

A naive interpretation of these VCA calculations would imply that the highest moment is obtained for Z=25.5, that is, substituting one Mn atom in every second unit cell could yield an enhanced moment/f.u.. The question of obtaining a ferromagnetically coupled Mn has previously been analyzed and has been found to be possible for surfaces \cite{ferromagMn1,ferromagMn2}, however there is no guarantee that a Mn and a Fe atom will have the same magnetic coupling as two identical VCA  atoms with Z=25.5.  In a further set of calculations, all three site II's were occupied  by identical VCA atoms. These showed a maximum in the calculated magnetic moment for  Z=25.85, that is the same number of electrons as in the previous  calculation with one VCA atom at site II with Z=25.5 and two regular Fe.   This indicates that the maximum in the ferromagnetic moment is controlled by  band filling, as required by the ideas behind the Slater-Pauling curve. However, when the VCA charge is reduced to Z=25.667, that is the number of electrons that in the previous set give a transition to an antiferromagnetic coupling (with one VCA atom at site 
II with Z=25.0 and two Fe) the predicted coupling continues to be ferromagnetic. This indicates that, unlike the maximum in the ferromagnetic moment, the transition from ferromagnetic to antiferromagnetic coupling is a local (atomic) property that is not well reproduced by VCA calculations. This is  further confirmed by a last set of calculations, with doubled and quadrupled cells with only one Mn atom which is found to couple antiferromagnetically in both cases. Finally, supercell calculations with the larger quadrupled structure, with 4 cells, and one substitutional atom (Ti, V or Cr) also show  that the stable magnetic solution is always with a ferrimagnetic coupling on the substitutional site and with no enhancement of the Fe moments, thus reducing the total moment as a whole.

\section{Conclusion}
We have carried out calculations to highlight the behavior of substitutional transition metal atoms  occupying either Fe I  or Fe II positions in Fe$_4$N perovskite crystals, thus giving a more robust frame to the  scattered results available so far in the literature. Motivated by the goal of  exploring   the best avenue for finding a giant magnetic moment in bulk symmetric iron-based ferromagnets, we are able to draw a coherent and complete picture of  magneto-volume effects in Fe$_4$N and to evaluate, on the ground of electronic structure calculations, the possibility of enhancing the magnetic properties by alloying with 3d metals.  

We have clearly shown by the results of  Figs.\ref{fig2}-\ref{fig4}  that the equilibrium value of the lattice parameter obtained by simulations must come out close to the experimental value, as found here,  in order to obtain a consistent picture of the ferromagnetic ground state.  In our calculations this is achieved by the use of GGA functional in both FP-LMTO and FPLO methods. The non-trivial behavior of Fe moments in Fe$_4$N was investigated several years ago by Mohn and Matar\cite{mohn-matar} who performed calculations only in LSDA, both by   Augmented Spherical Waves and FP-Linear Augmented Plane Waves methods, finding Fe I and Fe II moments in fair agreement with our LDA curve in Fig.\ref{fig3}.  Their results  confirm that this approximation  is   unable  to account for the anomalous behavior of Fe II moments, as also found in older calculations\cite{matar2}  adopting the experimental lattice parameter\cite{frazer} from the start. 

There are unfortunately very few measurements of the magnetization and local moments in these compounds to compare with theoretical results. Diffraction experiments by Frazer\cite{frazer} give a total moment of 9.0 $\mu_B$ f.u.  deduced from  3.0-2.98  $\mu_B$  f.u. and 2.0-2.1 $\mu_B$ for Fe I and Fe II atoms, respectively, whereas magnetization results with  8.86 $\mu_B$ f.u. are found by  Wiener and Berger\cite{momentfu1} and 11.6  $\mu_B$  by  Atiq {\it et al.}\cite{momentfu2}  using  Fe$_4$N thin films on substrates, which probably provides enhanced magnetic moments. 
The above experimental findings  are to be compared with our  9.93 (9.84) $\mu_B$  per unit cell, with atomic moments of 3.0 (2.91) and 2.30 (2.31) $\mu_B$, for the  Fe I and  Fe II atom, respectively, obtained from the FPLO (FP-LMTO) method using GGA. We notice that in all our calculations for Fe$_4$N the GGA calculated moments were larger than the LDA ones by about 1.5\% and 13\%-15\%, for FeI and FeII moment, respectively. As a whole, these considerations cannot neglect the magnetic behavior of parent bcc $\alpha$-Fe and fcc $\gamma$Fe,  weak (d-states partially quenched) and strong (d-states fully occupied ) ferromagnet, respectively, with 2.2 $\mu_B$/atom and 2.7$\mu_B$/atom. By simply taking the average value of 2.45 $\mu_B$/atom as  reference value, we conclude that our GGA calculations and the measurements by Frazer\cite{frazer} are sound. According to the same  criterion, the LDA results are ruled out. 
We therefore conclude that the use of GGA, in connection with these calculations, give a coherent picture of cohesive and magnetic properties of Fe$_4$N  describing the approach to magneto-volumic equilibrium both in the high and low spin regime. 

As shown in Figs.\ref{fig5}, \ref{fig6} substitution of Fe for other transition metals like Mn, Co or Ni, shows an intricate behavior in which the Mn substitution clearly favors the Fe II site, whereas Ni favors substitution on the Fe I site  as well as Co to a minor extent. Ni and Co substitution results in a ferromagnetic coupling to the Fe atoms, whereas Mn couples antiferromagnetically on the strongly preferential Fe II site and ferromagnetically on the Fe I site. Of all types of doping investigated here, only the energetically very unfavorable case of Mn doping at the Fe I site increases the magnetic moment. We thus conclude that enhancing the saturation moment of Fe$_4$N by simply exchanging one in four Fe atoms in the unit cell is not possible. However, the strong magneto-volume effects in Fe$_4$N, as well as in the Mn-doped system and previously indicated in the literature\cite{santos,patwari01prb64:214417}, means that there might be possibilities to maximize the magnetic moment per f.u., possibly in combination with exploring the effect of N vacancies and the effect of band filling on the type of magnetic coupling. It should also be pointed out that the present study has not considered the effect of disorder, which has previously found to have strong influence on the magnetic coupling in a related Fe-Mn system\cite{hudl2011}, and so it may still be possible to achieve sufficient ferromagnetism in a Mn doped system to enhance the saturation moment. Fe$_4$N  has  a saturation magnetization which is lower that that given by the Slater-Pauling maximum of bcc Fe-Co alloys. However, for practical applications the low cost of the elements constituting Fe$_4$N, could make it a competitive material whereas magneto-volume effects make the exploration of alloying with Co and Mn quite worthwhile for further investigation.

Our present findings,  push us  toward  a different  route for enhancing magnetism in iron and iron-doped nitrides: the  implantation of  Mn or  Co at site I or even II in layered compounds.  In fact, as indicated by  preliminary results by some  of us for  Fe$_4$N(001), the effect of being near a surface  enhances  significantly the magnetic moments both at   Fe I and Fe II not only on the top layer but also in farther  sublayers. The enhancing effect of layered structures is also found in Ref. 30.

\section{Acknowledgements}
O.E. gratefully acknowledges support from VR, the KAW foundation and the ERC (project 247062 - ASD). T.B. acknowledges support from the Academy of Finland centre of excellence program (COMP) and computational resources from Finlands IT-centre for Science (CSC). P.M. acknowledges fruitful discussions with colleagues of the Condensed Matter group during her LLP Erasmus TS visit at the CFMC-Universidade de Lisboa as well as  the hospitality and support of the Uppsala University, Department of Physics and Astronomy. The CINECA award under the ISCRA initiative for the availability of high performance computing resources and support is also acknowledged. T.G. thanks the FCT for funding through the project PEst-OE/FIS/UI0261/2011.


\end{document}